\documentclass[conference]{IEEEtran}
\usepackage[hidelinks]{hyperref}
\usepackage{tikz}
\usepackage[first=0,last=9]{lcg}
\usetikzlibrary{calc}
\usepackage{multicol}
\usepackage{ctable}
\usepackage{mdframed}
\usepackage{color, colortbl}
\usepackage{epsfig,subcaption}
\usepackage{caption}
\usepackage{amssymb, amsmath}
\usepackage{scrextend}
\usepackage{color}
\usepackage{bm}
\usepackage{amsthm}
\usepackage{cite}
\usepackage{pgfplots}
\usepackage{arydshln}

\usetikzlibrary{shapes,decorations,arrows,calc,arrows.meta,fit,positioning}

\usepackage{graphicx}
\usepackage{color}

\usepackage{amssymb}
\usepackage{amsmath,amsfonts,amssymb}
\usepackage{verbatim}
\usepackage{stfloats}
\usepackage[bookmarks=false]{}

\newtheorem{theorem}{Theorem}

\newtheorem{remark}{Remark}

\allowdisplaybreaks

\DeclareCaptionLabelFormat{adja-page}{\hrulefill\\#1 #2 \emph{(previous page)}}

\usetikzlibrary{decorations.markings}

\tikzstyle{interrupt}=[
postaction={
	decorate,
	decoration={markings,
		mark= at position 0.25
		with
		{
			\begin{scope}[-]
				\fill[white] (0,0) rectangle (0,0);
				\draw (-0.15,0.25) -- (0.15,-0.25);
				\draw (0.15,0.25) -- (-0.15,-0.25);
			\end{scope}
		}
	}
}
]

\begin{document}
	\title{Price of Precision in Coded Distributed Matrix Multiplication: A Dimensional Analysis}
	\author{\IEEEauthorblockN{Junge Wang, Zhuqing Jia and Syed A. Jafar}
		\IEEEauthorblockA{Center for Pervasive Communications and Computing (CPCC)\\
			University of California, Irvine \\
			Email: \{jungew, zhuqingj, syed\}@uci.edu}}
	\maketitle
	\begin{abstract}
		Coded distributed matrix multiplication (CDMM) schemes, such as MatDot codes, seek efficient ways to distribute  matrix multiplication task(s) to a set of $N$ distributed servers so that the answers returned from any $R$ servers are sufficient to recover the desired product(s). For example, to compute the product of matrices ${\bf U, V}$, MatDot codes partition each matrix into $p>1$ sub-matrices to create smaller coded computation tasks that reduce the upload/storage at each server by $1/p$, such that ${\bf UV}$ can be recovered from the answers returned by any $R=2p-1$ servers. An important concern in CDMM  is to reduce the recovery threshold $R$ for a given storage/upload constraint. Recently, Jeong et al. introduced Approximate MatDot (AMD) codes that are shown to improve the recovery threshold by a factor of nearly $2$, from $2p-1$ to $p$. A key observation that motivates our work is that the storage/upload required for approximate computing depends not only on the dimensions of the (coded) sub-matrices that are assigned to each server, but also on their precision levels --- a critical aspect that is not  explored by Jeong et al. Our main contribution is a rudimentary  dimensional analysis of AMD codes inspired by the Generalized Degrees of Freedom (GDoF) framework previously developed for wireless networks, which indicates that for the same upload/storage, once the precision levels of the task assignments are accounted for, AMD codes surprisingly fall short in all aspects of even the trivial replication scheme which assigns the full computation task to every server. Indeed, the trivial replication scheme has a much better recovery threshold of $1$, better download cost,  better computation cost, and much better encoding/decoding (none required) complexity  than AMD codes.  The dimensional analysis is supported by simple numerical experiments.
		
	\end{abstract}
	
	\section{Introduction}
\begin{figure}[!h]
		\centering
		\begin{tikzpicture}[xscale=0.6,yscale=0.7]
			\node[rectangle, help lines, fill=black!5, text=black, draw=black, minimum size=0.7cm, inner sep=0.2cm, rounded corners=0.5em] (S1) at (2cm, 1cm) { ${\mathbf{U}}$};
			
			\node[rectangle, draw=black, fill=black!5, text=black,minimum size=0.7cm,  inner sep=0.2cm, rounded corners=0.5em] (Si) at (14cm, 1cm) { ${\mathbf{V}}$};
			\node [draw, rectangle,fill=black!5, text=black, inner sep =0.2cm] (D1) at (2cm, -1.5cm) {\footnotesize Server $1$};
			\node [rectangle, inner sep =0.2cm] (Ddots1) at (4cm, -1.5cm) {$\cdots$};
			\node [draw, rectangle, fill=black!5, text=black, inner sep =0.2cm] (Dm) at (6cm, -1.5cm) {\footnotesize Server $i$};
			\node [rectangle, inner sep =0.2cm] (Ddots2) at (8cm, -1.5cm) {$\cdots$};
			\node [draw, rectangle, fill=black!5, text=black, inner sep =0.2cm] (Dn) at (10cm, -1.5cm) {\footnotesize Server $j$};
			\node [rectangle, inner sep =0.2cm] (Ddots2) at (12cm, -1.5cm) {$\cdots$};
			\node [draw, rectangle, fill=black!5, text=black, inner sep =0.2cm] (DN) at (14cm, -1.5cm) {\footnotesize Server $N$};
			
			\draw [violet, thick,  ->] (S1)--(DN) node[draw, rectangle, fill=white, pos=0.3]{\scriptsize ${\bf F}_N$};
			\draw [violet, thick,  ->] (S1)--(D1) node[draw, rectangle, fill=white, pos=0.3]{\scriptsize ${\bf F}_1$};
			\draw [violet, thick,  ->] (S1)--(Dn) node[draw, rectangle, fill=white, pos=0.3]{\scriptsize ${\bf F}_j$};
			\draw [violet, thick,  ->] (S1)--(Dm) node[draw, rectangle, fill=white, pos=0.3]{\scriptsize ${\bf F}_i$};
			
			\draw [blue, thick, ->] (Si)--(D1) node[draw, rectangle, fill=white, pos=0.3]{\scriptsize ${\bf G}_1$};
			\draw [blue, thick, ->] (Si)--(Dm) node[draw, rectangle, fill=white, pos=0.3]{\scriptsize ${\bf G}_i$};
			\draw [blue, thick, ->] (Si)--(Dn) node[draw, rectangle, fill=white, pos=0.3]{\scriptsize ${\bf G}_j$};
			\draw [blue, thick, ->] (Si)--(DN) node[draw, rectangle, fill=white, pos=0.3]{\scriptsize ${\bf G}_N$};

			\node[circle, draw=black, fill=black!5, text=black, minimum size=0.9cm, inner sep=0] (Uj) at (8cm, -4.5cm) { \small User};

			\draw [red, thick, ->] (D1)--(Uj) node[draw, rectangle, fill=white, pos=0.25]{\scriptsize $\mathbf{Y}_1$};
			\draw [red, thick, ->] (Dn)--(Uj) node[draw, rectangle, fill=white, pos=0.25]{\scriptsize $\mathbf{Y}_j$};
			\draw [red, thick, interrupt, ->] (Dm)--(Uj);
			\draw [red, thick, ->] (DN)--(Uj) node[draw, rectangle, fill=white, pos=0.25]{\scriptsize $\mathbf{Y}_N$};
			\node[right=0.5cm of Uj, minimum size=0.3cm, inner sep=0.1cm] (P) {\scriptsize  $\mathbf{U}\mathbf{V}$};
			\draw[->](Uj)--(P);
			\node [rectangle, rounded corners=0.5em, inner sep =0.2cm, fill=yellow, fill opacity=0.75, text opacity=1] at (8cm, -3.25cm) {A total of $R$ answers downloaded};
			
		\end{tikzpicture}
		\caption{\it  \small Coded Distributed Matrix Multiplication (CDMM). \vspace{-0.2cm}}
		\label{fig:CDMM}
	\end{figure}
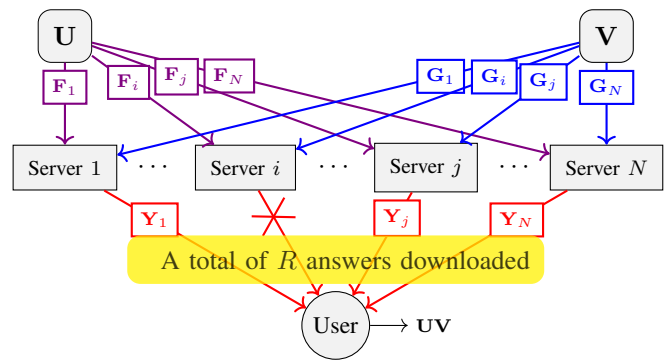

	Coded distributed matrix multiplication (CDMM) \cite{Yu_Maddah-Ali_Avestimehr_Polynomial,Dutta_Fahim_Haddadpour,GPolyDot, Yu_Maddah-Ali_Avestimehr, Yu_Lagrange, Reisizadeh_Prakash_Pedarsani, Lee_Suh_Ramchandran, Lee_Lam_Pedarsani, Dutta_Cadambe_Short, Dutta_Cadambe_Codedconv, Yu_Maddah-Ali_CodedDFT, Jahani-Nezhad_Maddah-Ali, Baharav_Lee_Ocal, Suh_Lee_Msparse, Wang_Liu_CLT, Mallick_Chaudhari_Joshi, Wang_Liu_Sparse, Severinson_iAmat_Rosnes, Haddadpour_Cadambe_Finite,Sheth_Dutta_Chaudhari, Jeong_Ye_Grover,Kim_Sohn_Moon_Group,Park_Lee_Sohn,Li_Maddah-Ali_Fog, Jia_Jafar_CDBC,Chen_Jia_Wang_Jafar}  (see Figure \ref{fig:CDMM}) seeks to distribute a matrix multiplication task among $N$ servers as efficiently as possible so that from the answers received from any $R$ responsive servers the sink (user) is able to recover the desired computation result. $R$ is referred to as {\it recovery threshold}. Existing state-of-the-art solutions \cite{Yu_Maddah-Ali_Avestimehr_Polynomial, Dutta_Fahim_Haddadpour,Yu_Maddah-Ali_Avestimehr} to CDMM are built upon matrix partitioning and polynomial based coding -- the constituent matrices $\mathbf{U}$, $\mathbf{V}$ are partitioned into block submatrices, coded shares of which are sent to the servers. The servers compute the products of their encoded shares, which can be viewed as evaluations of carefully constructed polynomials with partitioned block matrices as coefficients. Categorized by different partitioning strategies, state-of-the-art approaches fall into three classes: Polynomial codes\cite{Yu_Maddah-Ali_Avestimehr_Polynomial} for row-by-column partitioning, MatDot codes\cite{Dutta_Fahim_Haddadpour} for column-by-row partitioning and Entangled Polynomial codes (EP codes)\cite{Yu_Maddah-Ali_Avestimehr} for arbitrary partitioning. 
	
	When CDMM schemes are utilized for computations over real numbers, numerical stability concerns become important \cite{Fahim_Cadambe,Ramamoorthy_Li,Subramaniam_random}. As noted in the literature \cite{Gautschi_Vdm, Pan_Vandermonde}, this is because real Vandermonde matrices (which are essential in decoding) are ill-conditioned, especially for large $R$. In other words, small error in answers (that is a natural result of quantization) yields large error in decoded computation results. To overcome this problem, a variety of techniques are developed, for example, Chebyshev polynomial based coding schemes \cite{Fahim_Cadambe}, circulant and rotation matrix embeddings \cite{Ramamoorthy_Li}, and random Khatri-Rao product codes \cite{Subramaniam_random}.

	Another important concern in CDMM literature has been to find ways to reduce the recovery threshold for a given storage/upload cost per server.  In this regard, a recent breakthrough is reported in \cite{Cadambe_AMatdot} which introduces approximated CDMM solutions under the polynomial based coding framework. Based on MatDot codes\footnote{As in \cite{Cadambe_AMatdot}, the results generalize to $\epsilon$-approximate EP codes as well.} and a set of sufficiently small evaluation points, \cite{Cadambe_AMatdot} shows that Approximate MatDot (AMD) codes achieve the recovery threshold of $R=p$ with bounded error $\epsilon$. Compared with MatDot codes where the achieved recovery threshold is $R=2p-1$, this is {\it ``nearly twice as efficient as exact multiplication''}. 
	
	Since a  repetition code scheme which replicates the full computation task at every server can trivially achieve the recovery threshold of $R=1$, it is crucial that the recovery threshold be optimized subject to constraints on the upload or storage cost. On the one hand, because AMD codes use the same matrix partitioning by a factor of $1/p$ as MatDot codes, one might  imagine that just like MatDot codes, AMD codes require a fraction $1/p$ of the storage/upload cost  (compared to repetition codes). But on the other hand, when approximate computation is involved, the storage/upload cost also depends strongly on the precision with which numerical values need to be represented. Our work is motivated by the goal of properly accounting for the scaling of upload/storage costs as a function of the desired precision of computation.
	
	To accomplish this goal we take an approach inspired by the Generalized Degrees of Freedom (GDoF) framework that has been extensively used \cite{Etkin_Tse_Wang, Arash_Bofeng_Jafar_BC} to study the approximate and robust capacity limits of wireless networks. Based on a similar framework, our rudimentary dimensional analysis reveals insights consistent with \cite{Cadambe_AMatdot} in terms of the required `small' size of evaluation points.\footnote{Our analysis can also be extended to `large' evaluation points, but since the conclusion remains pessimistic, that case is omitted here.} Surprisingly, it also reveals a strongly pessimistic outlook of AMD codes. The advantage of reduced recovery threshold that is achieved by AMD codes, is shown to come at the cost of increased upload/storage costs by a factor of $p$, due to increased precision requirement from the uploads. Intuitively, this is because the small evaluation points of AMD codes produce extremely ill-conditioned decoding Vandermonde matrices. In fact the dimensional analysis shows that AMD codes give away so much in their upload/storage costs that they fall short  of even replication codes in all regards. Indeed, the trivial replication scheme has a much better recovery threshold of $1$, better download cost,  better computation cost, and much better encoding/decoding (none required) complexity  than AMD codes. The dimensional analysis is supported by modest numerical experiments.
		
	
	{\it Notation: }For  integers $m,n$ such that $m<n$, $[m:n]\triangleq\{m,m+1,\cdots,n\}$, $\mathbf{X}_{m:n}\triangleq\{\mathbf{X}_m,\mathbf{X}_{m+1},\cdots,\mathbf{X}_n\}$. $[n]\triangleq[1:n]$. $||.||_F$ denotes the Frobenius norm.  The notation $\widetilde{O}(a\log b)$ suppresses\footnote{There is another standard definition of the notation $\widetilde{O}$ which fully suppresses polylog terms, i.e, $O(a\text{polylog}(b))$ is represented by $\widetilde{O}(a)$. The definition used in this paper emphasizes the dominant factor in the polylog term.} 
	polylog terms. When an $m\times p$ matrix $\mathbf{A}$ is multiplied by a $p\lambda\times n\kappa$ matrix $\mathbf{B}$ where $\mathbf{B}$ is written as a block matrix with $p\times n$ blocks and the blocks have the dimension $\lambda\times\kappa$, the product $\mathbf{AB}$ means that $(\mathbf{A}\otimes\mathbf{I}_{\lambda})\mathbf{B}$ where $\otimes$ is the Kronecker product and $\mathbf{I}_{\lambda}$ is the $\lambda\times\lambda$ identity matrix.
	
	\section{Preliminaries}
	\subsection{MatDot Codes\cite{Dutta_Fahim_Haddadpour}}
	Partition matrices $\mathbf{U},\mathbf{V}$  as follows,
	\begin{align}
		\mathbf{U}=[\mathbf{U}_1,\mathbf{U}_2,\cdots,\mathbf{U}_p],\mathbf{V}=[\mathbf{V}_1,\mathbf{V}_2,\cdots,\mathbf{V}_p]^T
	\end{align}
	Let $\alpha_1,\alpha_2,\cdots,\alpha_N$ be distinct elements in $\mathbb{R}$. Server $i$, upon receiving the encoded matrices,
	\begin{align}
		{\bf F}_i&=\mathbf{U}_1+\alpha_i\mathbf{U}_2+\cdots+\alpha_i^{p-1}\mathbf{U}_p\\
		{\bf G}_i&=\mathbf{V}_p+\alpha_i\mathbf{V}_{p-1}+\cdots+\alpha_i^{p-1}\mathbf{V}_1,
	\end{align}
	computes the product,
	\begin{align}
		\mathbf{C}_i={\bf F}_i{\bf G}_i=\sum_{i=1}^p\sum_{j=1}^p \mathbf{U}_i\mathbf{V}_j \alpha_i^{p-1+i-j}.
	\end{align}
	The product can be viewed as a polynomial (with respect to $\alpha$) of degree $2p-2$. With the answers from any $R=2p-1$ servers, the user is able to recover all the coefficients of the polynomial, so that the desired matrix product $\mathbf{C}=\mathbf{U}\mathbf{V}=\sum_{i=1}^p \mathbf{U}_i\mathbf{V}_i$ can be recovered since it is the coefficient of the term $\alpha^{p-1}$. The recovery threshold is $R(p,0)=2p-1$. The notation is explained in the next section.
	\subsection{Approximate Computation \cite{Cadambe_AMatdot} }
	Assume $||\mathbf{U}||_F\leq \eta,||\mathbf{V}||_F\leq \eta$. The goal is to perform approximate computation so that
	\begin{align}\label{eq:approxcons}
		|\hat{\mathbf{C}}_{i,j}-\mathbf{C}_{i,j}|\leq \epsilon
	\end{align}
	The recovery threshold subject to the constraint \eqref{eq:approxcons}, is denoted as $R(p,\epsilon)$.	\cite{Cadambe_AMatdot} introduces  Approximate MatDot codes that are able to achieve the `\emph{optimal}' recovery threshold $R(p,\epsilon)=p$. The main result of \cite{Cadambe_AMatdot} is summarized below.
	{\begin{theorem}\cite{Cadambe_AMatdot} \label{thm:AMD}
			(Achievability) For any $0\leq\epsilon\leq \min(2,3\eta^2\sqrt{2p-1})$, AMD codes can be constructed by choosing the evaluation points $\alpha_1,\alpha_2,\cdots,\alpha_N$ as 
			\begin{align}
				|\alpha_i|\leq \frac{\epsilon}{6\eta^2\sqrt{2p-1}(p^2-p)} \label{eq:alphacondition}
			\end{align}
			such that  $R(p,\epsilon)=p$.
			
			(Converse) For all $0\leq\epsilon\leq\eta^2$, $R(p,\epsilon)\geq p$
		\end{theorem}
	}
	
		{\remark Intuitively, the key idea of AMD codes is to assign small values to the evaluation points $\alpha_i$, so that the terms with sufficiently high powers of $\alpha_i$ become essentially negligible, thus reducing the number of unknowns, which in turn reduces the recovery threshold.}
	
	\section{The Price of Precision in CDMM}

	\subsection{GDoF Framework}
	Inspired by the GDoF framework that has been used extensively \cite{Etkin_Tse_Wang, Arash_Bofeng_Jafar_BC}  to study the approximate capacity of wireless networks let us introduce a similar basic framework to study the fundamental tradeoffs in the approximate computing problem. As a toy example to introduce basic notation, using base-$B$ representation\footnote{GDoF studies \cite{Arash_Bofeng_Jafar_BC} of wireless networks use base $P$ representation, and $P$ is labeled `power' as a legacy  from prior DoF studies where it indeed represents transmit power.} for numerical values, consider a random variable that takes values, say in $[0,B^\mu)$, $B=10, \mu=4$, and let $W$ be its representation accurate to $\nu=5$ digit precision. $W$ may be represented as (cf. \cite{Arash_Bofeng_Jafar_BC}),
	\begin{align}
		W&=B^{\mu}(\overline{W} + B^{-\nu}\tilde{W})
	\end{align}
	with $\overline{W}, \tilde{W}\in[0,1)$, such that $B^\mu \overline{W}$ represents the actual value (with infinite precision),  $B^{-\nu}\tilde{W}$ represents noise that primarily affects (truncates or subtracts)  the digits that appear after the $\nu$ digits that are accurately known, and $W$ represents the truncated value.  For example, ${\bf w_1w_2w_3w_4.w_5} = 10^4(0.{\bf w_1w_2w_3w_4w_5}w_6w_7\cdots+10^{-5}(-0.w_6w_7\cdots))$, is a $5$ digit precision representation of the actual value ${\bf w_1w_2w_3w_4.w_5}{w_6w_7\cdots}$ where the precise digits are highlighted in bold. Note that the noise term $B^{-\nu}\tilde{W}$ is not independent of $\overline{W}$. Equivalently, one may view $W$ as the truncated version of $\overline{W}$, carrying only the  $\nu$ digits of $\overline{W}$ that are accurately known, i.e.,
	\begin{align}
		W&=B^{\mu-\nu}(\overline{W})^\nu,
	\end{align}
	where we define the notation,
	\begin{align}
		(\overline{W})^\nu\triangleq\lfloor B^\nu \overline{W}\rfloor,
	\end{align}
	to represent the integer value comprised of the top $\nu$ digits of the normalized quantity $\overline{W}$.
	This will be the standard notation throughout this work.
	
	The GDoF framework uses such exponential representations in a generic $B$-ary alphabet, and allows the base $B$  to approach infinity. This has the advantage that it removes lower order effects (because of normalization by $\log_2(B)$ when measuring information in $B$-ary units), thus smoothing out the finer details, e.g., the choice of the particular input distribution, which do not scale with alphabet size, and exposes the sharp fundamental tradeoffs that are typically expected from dimensional analysis. In fact, for the GDoF framework, instead of the interval $[0,1)$ it suffices to assume that $\overline{W}$ and $\tilde{W}$ are $O(1)$, i.e., bounded by some constants independent of $B$.  We will use the overline and tilde notations  throughout this work to represent the normalized (limited to $O(1)$) precise values and noise, and $\mu_\bullet,\nu_\bullet$ for the magnitude and precision levels, respectively. 


	\subsection{Dimensional Analysis: $\min(\nu_f,\nu_g)\geq p\nu$}
	As the main result of this work we will show in this section that for AMD codes  the uploads to each server (${\bf F}_i, {\bf G}_i$) need precision levels ($\nu_f, \nu_g$) that are at least $p$ times greater than the precision level of the computed product ($\nu$). In other words, the main result is the bound, $$\min(\nu_f,\nu_g)\geq p\nu.$$
	
	Using notation consistent with the GDoF  formulation, the uploads to the servers are represented as
	\begin{align}
		{\bf F}_i&=B^{\mu_f}(\overline{\bf F}_i+B^{-\nu_f}\tilde{\bf F}_i) = B^{\mu_f-\nu_f}(\overline{\bf F}_i)^{\nu_f}\\
		{\bf G}_i&=B^{\mu_g}(\overline{\bf G}_i+B^{-\nu_g}\tilde{\bf G}_i)=B^{\mu_g-\nu_g}(\overline{\bf G}_i)^{\nu_g}\\
		\intertext{where}
		\overline{\bf F}_i&={\bf \overline{U}}_1+\alpha_i \overline{\bf U}_2+\cdots+\alpha_i^{p-1} \overline{\bf U}_p\\
		\overline{\bf G}_i&=\alpha_i^{p-1}\overline{\bf V}_1+\alpha_i^{p-2} \overline{\bf V}_2+\cdots+ \overline{\bf V}_p.
	\end{align}
	It is assumed that the normalized inputs $\overline{\bf U}_i,\overline{\bf V}_i$, and the noise terms $\tilde{\bf F}_i, \tilde{\bf G}_i$ are all $O(1)$. Thus, the uploads ${\bf F}_i, {\bf G}_i$ have precision $\nu_f, \nu_g$, respectively. The scaling factors $B^{\mu_f}, B^{\mu_g}$ can be normalized away in this setting, but let us keep them for an explicit representation of the scale of ${\bf F}_i$ and ${\bf G}_i$. Following the insights of \cite{Cadambe_AMatdot}, the distinct constants $\alpha_i$ are assumed to be small, say 
	\begin{align}
		\alpha_i=B^{-\delta}\overline{\alpha_i}
	\end{align}
	for some $\delta> 0$ and $\overline{\alpha_i}=\Theta(1)$. The products ${\bf F}_i{\bf G}_i$ computed by the servers are expressed as follows.
	\begin{align}
		{\bf F}_i{\bf G}_i&=B^{\mu_f+\mu_g}\left(\overline{\bf F}_i\overline{\bf G}_i+B^{-\nu_f}\tilde{\bf F}_i\overline{\bf G}_i+B^{-\nu_g}\overline{\bf F}_i\tilde{\bf G}_i\right.\notag\\
		&\hspace{2cm}\left.+B^{-\nu_f-\nu_g}\tilde{\bf F}_i\tilde{\bf G}_i\right)
	\end{align}
	which has precision limited  to $\min(\nu_f,\nu_g)$ digits because of the additional noise terms. From  Server $i$, the user downloads ${\bf F}_i{\bf G}_i$ to $\nu_y$ digit precision. Since ${\bf F}_i{\bf G}_i$ has only $\min(\nu_f,\nu_g)$ digit precision, we require
	\begin{align}
		\nu_y\leq\min(\nu_f,\nu_g).
	\end{align}
	The download from Server $i$ is then represented as
	\begin{align}
		{\bf Y}_i &=B^{\mu_f+\mu_g}(\overline{\bf Y}_i+B^{-\nu_y}\tilde{\bf Y}_i) = B^{\mu_f+\mu_g-\nu_y}(\overline{\bf Y}_i)^{\nu_y}\\
		\overline{\bf Y}_i&=\overline{\bf F}_i\overline{\bf G}_i\\
		&=\overline{\bf U}_1\overline{\bf V}_{p}+\alpha_i (\overline{\bf U}_1\overline{\bf V}_{p-1}+\overline{\bf U}_{p-1}\overline{\bf V}_p)+\cdots\notag\\
		&+\alpha_i^{p-1}(\overline{\bf U}_1\overline{\bf V}_1+\cdots+\overline{\bf U}_p\overline{\bf V}_p)+\alpha_i^{2p-2}(\overline{\bf U}_p\overline{\bf V}_1)\\
		&=\overline{\bf X}_0+\alpha_i \overline{\bf X}_1+\cdots+\alpha_i^{2p-2}\overline{\bf X}_{2p-2}
	\end{align}
	The compact notation $\overline{\bf X}_i$ is used for the $O(1)$ terms that represent the corresponding sums of various $\overline{\bf U}_i\overline{\bf V}_j$ terms. The desired term is $\overline{\bf X}_{p-1}$.
	
	A recovery threshold of $p$ means that decoding must be accomplished with only $p$ server responses. Without loss of generality, say we have the responses ${\bf Y}_1, \cdots, {\bf Y}_p$. 
		
	\begin{align}
		&\left[\begin{matrix}{\bf Y}_1 \\ {\bf Y}_2\\ \vdots\\ {\bf Y}_{p}\end{matrix}\right]
		=B^{\mu_f+\mu_g}\left[
		{\arraycolsep=2pt\begin{matrix}
				1 & \alpha_1 & \cdots &\alpha_1^{p-1}\\
				1& \alpha_2 & \cdots &\alpha_2^{p-1}\\
				\vdots&\vdots&\cdots&\vdots\\
				1& \alpha_p  &\cdots &\alpha_p^{p-1}
		\end{matrix}}
		\right]\left[\begin{matrix}\overline{\bf X}_0 \\ \overline{\bf X}_1\\ \vdots\\ \overline{\bf X}_{p-1}\end{matrix}\right]\notag\\
		&+
		B^{\mu_f+\mu_g}\left[
		{\arraycolsep=2pt\begin{matrix}
				\alpha_1^p & \cdots &\alpha_1^{2p-2}\\
				\alpha_2^p & \cdots &\alpha_2^{2p-2}\\
				\vdots&\cdots&\vdots\\
				\alpha_p^p  &\cdots &\alpha_p^{2p-2}
		\end{matrix}}
		\right]\left[\begin{matrix}\overline{\bf X}_p \\ \overline{\bf X}_{p+1}\\ \vdots\\ \overline{\bf X}_{2p-2}\end{matrix}\right]+B^{\mu_f+\mu_g-\nu_y}\left[\begin{matrix}{\bf \tilde{Y}}_1 \\ {\bf \tilde{Y}}_2\\ \vdots\\ {\bf \tilde{Y}}_{p}\end{matrix}\right]
	\end{align}
	From here, with certain (`mild') additional assumptions (see Section \ref{sec:conc} and Section \ref{sec:app}) it can be information theoretically argued  that the desired quantity $\overline{\bf X}_{p-1}=\overline{\bf U}\overline{\bf V}$ can  be recovered from ${\bf Y}$  with precision no higher than $(\min(\nu_y,p\delta)-(p-1)\delta)^+$. We defer the information theoretic derivation to Section \ref{sec:app} at the end of this paper, and provide here an intuitive justification instead, as follows.

	Solving for $\overline{\bf X}_{p-1}=\overline{\bf U}\overline{\bf V}$ involves an inversion of the first Vandermonde matrix while treating the other terms as noise. The first Vandermonde matrix has condition number at least $\Omega(B^{(p-1)\delta})$, which causes a noise amplification by that factor in the remaining terms. Intuitively, since $\overline{\bf X}_0,\overline{\bf X}_1, \cdots, \overline{\bf X}_{p-2}$ dominate $\overline{\bf X}_{p-1}$, this amounts to projection of the ${\bf Y}$ vector along the vector that lies in the null space of the first $p-1$ columns of the first vandermonde matrix. In this projected dimension, ${\bf X}_{p-1}$ can be recovered as the dominant term, and the remaining noise level is determined by the stronger of the two projected noise terms: the projection of $\tilde{\bf Y}$, which has strength $B^{\mu_f+\mu_g+(p-1)\delta-\nu_y}$, and the projection of $\overline{\bf X}_p$, which has strength $B^{\mu_f+\mu_g+(p-1)\delta-p\delta}$. Note that the $B^{(p-1)\delta}$ scaling factor appears in each case due to the noise amplification impact of the inversion of the first Vandermonde matrix. This allows the user to recover
	\begin{align}
		{\bf UV}&=\overline{\bf UV}+B^{(p-1)\delta-\min(\nu_y,p\delta)}\widetilde{\bf UV}
	\end{align}
	Thus, the answer can be recovered with $\nu$ digit precision, provided that,
	\begin{align}
		& \nu\leq \min(\nu_y,p\delta)-(p-1)\delta\\
		&\leq \min(\nu_f,\nu_g,p\delta)-(p-1)\delta\\
		&\implies \left\{
		\begin{array}{ll}
			\nu&\leq \nu_f-(p-1)\delta\\
			\nu&\leq \nu_g-(p-1)\delta\\
			\nu&\leq \delta
		\end{array}\right.\label{eq:yield}
	\end{align}
	Thus, the dimensional analysis  yields a bound on the required precision of the uploads as,
	\begin{align}
		\nu_f&\geq \nu+(p-1)\delta\\
		&\geq\nu+(p-1)\nu\\
		&=p\nu
	\end{align}
	Similarly, $\nu_g\geq p\nu$. 

\section{Observations}
 Let us interpret the result of \cite{Cadambe_AMatdot} that is summarized as Theorem \ref{thm:AMD} in this paper, in GDoF terms. To this end, let $\eta=B^{\mu/2},\epsilon=B^{\mu-\nu}$, where $\nu$ can be regarded as the precision level. Thus the entries of $\mathbf{U},\mathbf{V}$ are of the order $B^{\mu/2}$, so that the entries of $\mathbf{C}$ are of the order $B^{\mu}$. The precision level of each entry is at least $\nu$ digits in the $B$-ary alphabet. \cite{Cadambe_AMatdot} shows that to evaluate the matrix product to $\nu$ digit precision, the choice of the $\alpha_i$ should satisfy  condition \eqref{eq:alphacondition} which is re-stated as follows in GDoF terms,
	\begin{align}
		|\alpha_i|=B^{-\delta}\overline{\alpha_i}&\leq \frac{B^{\mu-\nu}}{6B^{\mu}\sqrt{2p-1}(p^2-p)}=O(B^{-\nu})\\
		\implies \delta &\geq \nu
	\end{align}
	This is indeed one of the conditions that we find from our dimensional analysis as well, as it appears in \eqref{eq:yield}.

The dimensional analysis goes a bit further, and reveals a rather pessimistic outlook according to which AMD codes fall short of even trivial  repetition codes in all aspects. A repetition code refers to the  scheme that  assigns the full computation task to each server by uploading the entire ${\bf U,V}$ matrices to each server, with $\nu$ digit precision, and downloads the result of the computation from any $1$ server, also to $\nu$ digit precision. On the other hand, AMD codes upload submatrices that are smaller by $1/p$ in terms of their number of elements, but with precision $p\nu$ for each element, which is $p$ times larger, so they have the same upload/storage cost as repetition codes. Moreover, in terms of recovery threshold, computation cost, and download cost, AMD codes are strictly worse than repetition codes.
	
	Repetition codes have a recovery threshold of $1$ because the download from any $1$ server suffices. While repetition codes download the answer ${\bf UV}$ as a $\lambda\times \lambda$ matrix to $\nu$ digit precision from only one server, AMD codes download a $\lambda\times\lambda$ matrix from each of $p$ servers, in each case to $p\nu$ precision, so the download cost of AMD codes is $p^2$ times greater than repetition codes. In terms of computation cost, recall that the complexity of multiplying two $n$ digit numbers is super-linear in $n$ --- the trivial multiplication scheme has complexity $O(n^2)$ but the Schönhage–-Strassen algorithm reduces it to $\widetilde{O}(n\log n)$ which is still superlinear. Now, AMD codes require fewer multiplications by a factor of $1/p$ due to matrix partitioning, however, since each multiplication is between numbers with a greater number of digits by a factor of $p$, and the complexity of multiplication is super linear in the number of digits, it turns out that AMD codes require greater computation complexity at each server, compared to repetition codes. Specifically, AMD codes require $p$ times fewer multiplications than repetition codes but each multiplication has complexity $\widetilde{O}(p\nu\log(p\nu))$ for AMD codes, as compared to $\widetilde{O}(\nu\log\nu)$ for repetition codes. The comparison is  illustrated in Table \ref{Table:comparison}. While encoding and decoding complexities are not listed in the table, note that  repetition codes do not require encoding/decoding at all, so AMD codes fall short of repetition codes in this regard as well. 	
	\begin{table}[!ht]
		\begin{tabular}{c|c|c|c}
			\hline
			& MatDot codes & AMD codes    & Repetition codes \\ \hline
			\begin{tabular}[c]{@{}c@{}}Recovery\\ threshold\end{tabular}  & $2p-1$      & $p$    & $1$              \\ \hline
			\begin{tabular}[c]{@{}c@{}}Upload/storage \\ per server\end{tabular}   & $ \nu/p$       & $\nu$ & $\nu$           \\ \hline
			\begin{tabular}[c]{@{}c@{}} Total\\ Download \end{tabular} & $(2p-1)\nu$       & $p^2\nu$ & $\nu$            \\ \hline
			\begin{tabular}[c]{@{}c@{}}Computation\\ cost per server\end{tabular} & $\widetilde{O}(\frac{\nu \log \nu}{p})$       & $\widetilde{O}(\nu \log (p\nu))$ & $\widetilde{O}(\nu \log \nu)$            \\ \hline
		\end{tabular}
		\caption{\it MatDot codes vs AMD codes vs repetition codes. Values shown are relative to each other. }
		\label{Table:comparison}
	\end{table}
	
	Last but not the least, it is important to also note the caveat that while dimensional analysis allows elegant characterizations of fundamental tradeoffs, this elegance relies on asymptotic analysis that neglects lower order effects. As such, in settings where the lower order effects are important, e.g., where matrices comprised of small numbers are being multiplied to low precision so the large $B$ assumption is not justified, it is conceivable that the conclusions of the dimensional analysis may be violated. While dimensional analysis  informs our intuition and provides  principled reasoning at a high level, ultimately numerical results are still important to fully reveal the finer tradeoffs for particular settings. Elaborate experiments are beyond the scope of this work, but modest numerical results are provided next, that indeed validate the insights from the dimensional analysis.
	
	\section{Numerical Results}
	Consider a simple setting where $p=3$, and the dimensions of the $\mathbf{U}$, $\mathbf{V}$ matrices are $1\times 3$ and $3\times 1$, respectively, $\mathbf{U}=[U_1, U_2, U_3], \mathbf{V}=[V_1, V_2, V_3]^T$, 	where $U_1, U_2, U_3, V_1, V_2, V_3$ are uniformly i.i.d. over $[0,1]$. We use base $B=10$, i.e., decimal representations.  The recovery threshold of AMD codes for this setting is $R=3$, and to simplify\footnote{Note that indeed, this is the best-case scenario for AMD codes. When there are stragglers, i.e., $N>R$, the condition number of the corresponding decoding matrix is even worse.} our simulation, we consider the setting $N=R=3$.  	The encoded version of the constituent matrices for Server $i$ are the following two scalars.
	\begin{align}
		F_i&=\text{truncate}(U_1+\alpha_iU_2+\alpha_i^2U_3,\gamma),\\
		G_i&=\text{truncate}(V_3+\alpha_iV_2+\alpha_i^2V_1,\gamma),
	\end{align}
	where the function $\text{truncate}(x,\gamma)$ truncates the value of $x$ at $\gamma$ digits after the decimal. The answer returned by Server $i$ is $Y_i=\text{truncate}(F_iG_i,\gamma)$. 	Denote $\max_{i\in[N]} \alpha_i$ as $\alpha_{\max}$, the selection of evaluation nodes $\alpha_1, \alpha_2, \alpha_3$ is given as $		\alpha_i = \frac{i}{N} \alpha_{\max}, \forall i\in[N]$. 	We use the decoding algorithm\footnote{To completely characterize the trade-off between upload/download costs, approximation error $\epsilon$ and the choice of $\alpha_{\max}$, we do not declare failure in our decoding algorithm even if the norm of the minimum norm solution exceeds the threshold $\sqrt{2p-1}\eta^2$ in Algorithm 1 of \cite{Cadambe_AMatdot}.}  \cite[(55)]{Cadambe_AMatdot}, i.e., the minimum norm solution to decode $\epsilon$-approximate MatDot codes. 
	
		Figure \ref{fig:pvu} plots the Monte Carlo simulation results of upload/download cost per server (i.e., $\gamma$) versus mean absolute error (MAE) for $\gamma\in\{4,5,\cdots,16\}$ and $\alpha_{\max}=10^{-4}$. It is evident that to achieve the desired approximation error of $10^{-4}$, i.e., $\nu=4$, the upload/download cost per server required is at least $\gamma=12=3\times 4=p\nu$, which confirms our analytical result. Figure \ref{fig:avp} plots Monte Carlo simulation results of $\alpha_{\max}$ versus MAE for $\alpha_{\max}\in\{10^{-7}, 10^{-6},\cdots, 10^{-1}\}$ and $\gamma=12$. Simulation results show that for the given upload/download cost per server $\gamma=12$, the best approximation error is achieved when $\alpha_{\max}=10^{-4}=10^{-12/3}=10^{-\gamma/p}$. Since  given $\gamma=12$, as illustrated in the red line in Figure \ref{fig:avp}, repetition codes (which do not depend on the selection of $\alpha_{\max}$) achieve the MAE of no more than $10^{-4}$ with the same upload cost, this again confirms our analytical results that $\epsilon$-approximate MatDot codes fall short of repetition codes.
	
	\begin{figure}
		\centering
		\begin{tikzpicture}[xscale=0.81,yscale=0.81]
			\begin{semilogyaxis}
				[xmin=4, xmax=16, ymin = 5e-5,grid style={line width=.1pt, draw=gray!20}, major grid style={line width=.2pt,draw=gray!50}, grid=both, axis lines = left, minor tick num=1,xlabel = {\large Upload cost per server, $\gamma$}, ylabel = {\large Mean absolute error},]
				\addplot [color=blue, ultra thick] table [x index = {0}, y index = {1}, col sep=comma] {1.csv};
			\end{semilogyaxis}
		\end{tikzpicture}\hspace{-0.5cm}
		\caption{Upload cost per server $\gamma$  vs mean absolute error.}	\label{fig:pvu}
		\begin{tikzpicture}[xscale=0.81,yscale=0.81]
			\begin{semilogyaxis}
				[xmin=-7, xmax=-1, ymin = 1e-5,grid style={line width=.1pt, draw=gray!20}, major grid style={line width=.2pt,draw=gray!50}, grid=both, axis lines = left, minor tick num=1,xlabel = {\large $\log_{10}(\alpha_{\max})$}, ylabel = {\large Mean absolute error},]
				\addplot [color=blue, ultra thick] table [x index = {0}, y index = {1}, col sep=comma] {2.csv};
				\addplot [color=red, ultra thick] coordinates {(-7,0.0001)
					(-1,0.0001)};
			\end{semilogyaxis}
		\end{tikzpicture}
		\caption{ $\log_{10}(\alpha_{\max})$ versus mean absolute error.}
		\label{fig:avp}
	\end{figure}

	\section{Conclusion}\label{sec:conc}
	The nature of our analysis is that of a converse argument, i.e., an impossibility result, which is only as strong as the generality with which it applies. So it is important to note its limitations. For instance,  the information theoretic analysis in Section \ref{sec:app} assumes $\overline{\bf X}_i$ are independent and scalars, but neither of those assumptions is beyond reproach. Indeed, while $\overline{\bf X}_i$ are perhaps algebraically independent, they may not be statistically independent, and in general they can certainly  be matrices. The assumptions of Section \ref{sec:app} are still meaningful, in that the converse applies to any scheme that does not take advantage of any potential dependence between $\overline{\bf X}_i$ and that decodes each element of the $\overline{\bf X}_{p-1}$ matrix by the same decoding rule. To our knowledge, this is true for all existing CDMM schemes, including AMD codes, and is likely to be true for most future schemes as well,  because the applications for CDMM typically require low decoding complexity. Similarly, aside from their magnitude constraints, the evaluation points $\alpha_i$ are assumed `generic', which is also true for all known schemes, but specialized choices of $\alpha_i$ that achieve alignment may be possible. This possibility is reminiscent of rational alignment in wireless networks \cite{Motahari_Gharan_Khandani_real}. On the other hand, even if such constructions are possible, the limited precision aspect may negate their benefits, if analogies may be drawn from wireless GDoF studies \cite{Arash_Jafar}. Nevertheless, all such limitations of current analysis leave the door open for future surprises. So while the benefits of AMD codes are indeed called in question by current analysis, we do not expect this pessimistic outlook to be the final word along this  new research avenue. On the contrary, we are optimistic that  the idea of exploiting the power dimension, that is introduced by AMD codes, may find clever uses in coded computing to enable new forms of interference alignment \cite{Cadambe_Jafar_int, Jafar_FnT}, just as the power dimension plays a critical role in the GDoF characterizations of wireless networks. Better formalizations of the GDoF perspective for distributed computing, that improve upon our rudimentary attempt in this work, may be the key to future advances along this promising research avenue.
	
	
	\section{Appendix}\label{sec:app}
	Let us  make the simplifying assumption that $\overline{\bf X}_i, i\in[0:2p-2]$ are independent scalars, and that there exists a finite  constant $\Delta$ such that the joint differential entropy of any non-empty subset $S\subset\{\overline{\bf X}_i: i\in[0:2p-2]\}$ is bounded as $h(S)>\Delta$ (bits). The assumption of independence of $\overline{\bf X}_i$ and that they are scalars may appear rather restrictive, because in practice $\overline{\bf X}_i$ may be neither, but the assumption is general enough to encompass any decoding scheme that does not exploit potential dependencies across $\overline{\bf X}_i$, and which applies the same decoding rule to recover each element of the $\overline{\bf X}_{p-1}=\overline{\bf UV}$ matrix.  See Section \ref{sec:conc} for additional discussion of such limitations.
	
	%
	
	  The $\nu$ digit precision of the recovered computed value is represented by the following  bound on the mean absolute error distortion,
	\begin{align}
	\mathbb{E}|\overline{\bf UV}-{\bf UV}|&=O(B^{-\nu}).\label{eq:mae}
	\end{align}
	Therefore, we have
	\begin{align}
	I(\overline{\bf UV}; {\bf UV})&=h(\overline{\bf UV}) - h(\overline{\bf UV}\mid {\bf UV})\\
	&\geq \Delta - h(\overline{\bf UV}-{\bf UV}\mid {\bf UV})\\
	&=\nu\log(B) + o(\log(B))\label{eq:boundLaplace}
	\end{align}
	 For  step \eqref{eq:boundLaplace} we used the fact that Laplace distributions are (differential) entropy maximizers subject to a mean absolute deviation\footnote{The choice of mean absolute error vs mean squared error in \eqref{eq:mae} is inconsequential from a GDoF perspective. For example, if the precision constraint \eqref{eq:mae} is framed instead in terms of mean squared error, i.e., MSE $=O(B^{-2\nu})$, the same GDoF bound is still obtained by using the fact that Gaussians are entropy maximizers subject to a variance constraint. } constraint. Now, if any decoding rule applied to ${\bf Y}_{1:p}$ recovers ${\bf UV}$ which represents $\overline{\bf X}_{p-1}=\overline{\bf UV}$ to $\nu$-digit precision, then we have the Markov Chain $\overline{\bf X}_{p-1}=\overline{\bf UV}\leftrightarrow {\bf Y}_{1:p}\leftrightarrow {\bf UV}$. From the GDoF perspective we  have,
	\begin{align}
		\nu&\leq \lim_{B\rightarrow\infty}\frac{I(\overline{\bf UV};{\bf UV})}{\log(B)}\\
 		&\leq \lim_{B\rightarrow\infty}\frac{I(\overline{\bf X}_{p-1};{\bf Y}_{1:p})}{\log(B)}\label{eq:GDoF}
	\end{align} 
	For cleaner notation we will occasionally suppress  $o(\log(B))$ terms that are inconsequential for GDoF according to \eqref{eq:GDoF}.
	
	\begin{align}
		{\small I(\overline{\bf X}_{p-1};{\bf Y}_{1:p})}&\leq I(\overline{\bf X}_{p-1};\overline{\bf Y}_{1:p})\\&=h(\overline{\bf Y}_{1:p})-h(\overline{\bf Y}_{1:p}\mid \overline{\bf X}_{p-1})\label{eq:MI}
	\end{align}
	\begin{align}
		&h(\overline{\bf Y}_{1:p})=h(\overline{\bf Y}_1)+h(\overline{\bf Y}_2\mid \overline{\bf Y}_1)+\cdots+h(\overline{\bf Y}_p\mid \overline{\bf Y}_{1:p-1})\notag\\
		&\leq -(0+1+\cdots+(p-1))\delta\log(B) +o(\log(B))\label{eq:B1}
	\end{align}
	This is because conditioning on $\overline{\bf Y}_{1:i-1}$ allows (Gaussian) elimination of $\overline{\bf X}_{0:i-2}$ terms from $\overline{\bf Y}_{i}$, leaving the dominant term  as $B^{-(i-1)\delta}\overline{\bf X}_{i-1}$ whose bounded support limits its  entropy to $-(i-1)\delta\log(B)$ (uniform distribution maximizes entropy). Next we bound the other entropy term.
	\begin{align}
		&h(\overline{\bf Y}_{1:p}\mid \overline{\bf X}_{p-1})\geq h(\overline{\bf Y}_{1:p}\mid \overline{\bf X}_{p-1}, \overline{\bf X}_{p+1:2p-2})\\
		&=h\left(\mathbf{Q}\left[\begin{matrix}\overline{\bf X}_0 \\\vdots\\ \overline{\bf X}_{p-2}\label{eq:105}\\ \overline{\bf X}_{p}\end{matrix}\right]\right)=h(\overline{\bf X}_{0:p-2},\overline{\bf X}_{p})+\log|\det(\mathbf{Q})|\\
		&\geq \Delta+\log|\det(\mathbf{Q})|=\log|\det(\mathbf{Q})| +o(\log(B)),\label{eq:106}
	\end{align}
	where 
	\begin{align}
		\mathbf{Q}=\left[\begin{matrix}
			1  & \cdots &\overline{\alpha}_1^{p-2}B^{-(p-2)\delta}&\overline{\alpha}_1^{p}B^{-p\delta}\\
			1  & \cdots &\overline{\alpha}_2^{p-2}B^{-(p-2)\delta}&\overline{\alpha}_2^{p}B^{-p\delta}\\
			\vdots&\vdots&\vdots&\vdots\\
			1  & \cdots &\overline{\alpha}_p^{p-2}B^{-(p-2)\delta}&\overline{\alpha}_p^{p}B^{-p\delta}
		\end{matrix}\right].
	\end{align}
	Note that  \eqref{eq:105} follows  by the assumption of independence. The determinant of $\mathbf{Q}$ can be approximated as
	\begin{align}
		|\det(\mathbf{Q})|&=O(B^{-(0+1+2+\cdots+(p-2)+p)\delta}).
	\end{align}
	Substituting  this approximation into \eqref{eq:106}, we have
	\begin{align}
		&h(\overline{\bf Y}_{1:p}\mid \overline{\bf X}_{p-1})\geq-(1+2+\cdots+(p-2)+p)\delta\log(B).\label{eq:107}
	\end{align}
	Combining \eqref{eq:GDoF}, \eqref{eq:MI}, \eqref{eq:B1} and \eqref{eq:107}, we have our first desired bound, 
	\begin{align}
	\nu&\leq \delta.
	\end{align}

		\begin{figure}[b]
		\centering
		\begin{tikzpicture}[xscale=1.55,yscale=0.85]
			\draw [blue,fill=cyan!30!white,very thick] (0,-3) rectangle (0.5,3)node [black,pos=.5,yshift=2cm] {\tiny $\overline{\bf X}_0$};
			
			\draw [draw=none,fill=black!20!white] (0.015,-3.02) rectangle (0.485,0.1);
			
			\draw[arrows=<->] (0.65,3)--(0.65,2.5)node[right,pos=0.5]{\small $\delta$};
			\draw[blue,fill=cyan!30!white,very thick] (0.5,-3) rectangle (1,2.5)node [black,pos=.5,yshift=2cm] {\tiny $\overline{\bf X}_1$};
			
			\draw [draw=none,fill=black!20!white] (0.515,-3.02) rectangle (0.985,0.1);
			
			\draw[draw=none] (1,0) rectangle (1.5,2.0)node [black,pos=.5] {\tiny $\cdots$};
			
			\draw[blue,fill=cyan!30!white,very thick] (1.5,-3) rectangle (2,1.1)node [black,pos=.5,yshift=1.5cm] {\tiny $\overline{\bf X}_{p-2}$};
			
			\draw [draw=none,fill=black!20!white] (1.515,-3.02) rectangle (1.985,0.1);
			
			\draw[arrows=<->] (2.15,1.1)--(2.15,0.6)node[right,pos=0.5]{\small $\delta$};
			\draw[blue,fill=cyan!30!white,very thick] (2,-3) rectangle (2.5,0.6)node [black,pos=.5,yshift=1.35cm] {\tiny $\overline{\bf X}_{p-1}$};
			
			\draw [draw=none,fill=black!20!white] (2.015,-3.02) rectangle (2.485,0.1);
			
			\draw[arrows=<->] (2.65,0.6)--(2.65,0.1)node[right,pos=0.5]{\small $\delta$};
			
			\draw [blue,fill=black!20!white,very thick] (2.5,0.1) rectangle (3,-3)node [black,pos=.5] {\tiny $\overline{\bf X}_{p}$};
			\draw[arrows=<->] (3.15,0.1)--(3.15,-0.4)node[right,pos=0.5]{\small $\delta$};
			\draw [draw=none,fill=black!20!white] (2.515,-3.02) rectangle (2.985,-2);
			\draw [blue,fill=black!20!white,very thick] (3,-0.4) rectangle (3.5,-3)node [black,pos=.5] {\tiny $\overline{\bf X}_{p+1}$};
			\draw [draw=none,fill=black!20!white] (3.015,-3.02) rectangle (3.485,-2.97); 
			\draw[draw=none] (3.5,-1.5) rectangle (4,-2)node [black,pos=.5] {\tiny $\cdots$};  
			\draw [blue,fill=black!20!white,very thick] (4,-2.0) rectangle (4.5,-3)node [black,pos=.5] {\tiny \scalebox{0.75}{$\overline{\bf X}_{2p-3}$}};
			\draw [draw=none,fill=black!20!white] (4.015,-3.02) rectangle (4.485,-2.97);  
			\draw[arrows=<->] (4.65,-2.5)--(4.65,-2)node[right,pos=0.5]{\small $\delta$};
			\draw [blue,fill=black!20!white,very thick] (4.5,-2.5) rectangle (5,-3)node [black,pos=.5] {\tiny \scalebox{0.75}{$\overline{\bf X}_{2p-2}$}};
			\draw [draw=none,fill=black!20!white] (4.515,-3.02) rectangle (4.985,-2.97); 
			
			\draw[ultra thick,rounded corners=0.25cm] (-0.5,0.1) rectangle (3.5,3)node[black, pos=0.5, xshift=1.25cm, yshift=0.5cm]{ $(\overline{\bf Y})^{\nu_y}$};
			\draw[arrows=<->] (-0.15,0.1)--(-0.15,3)node[left,pos=0.5]{\small $\nu_y$};
		\end{tikzpicture}
		\caption{The digit levels where various $\overline{\bf X}_i$ appear in ${\bf Y}$}
		\label{fig:gdof}
	\end{figure}

	Our next bound is obtained as follows.
	\begin{align}
		I(\overline{\bf X}_{p-1};{\bf Y}_{1:p})&=I(\overline{\bf X}_{p-1};(\overline{\bf Y}_{1:p})^{\nu_y})\\
		&\leq I\left(\overline{\bf X}_{p-1}; \left\{(\overline{\bf X}_i)^{(\nu_y-i\delta)^+}\right\}_{i\in[0:p-1]}\right)\label{eq:figjustify}\\
		&=I\left(\overline{\bf X}_{p-1}; (\overline{\bf X}_{p-1})^{(\nu_y-(p-1)\delta)^+}\right)\\
		&\leq H((\overline{\bf X}_{p-1})^{(\nu_y-(p-1)\delta)^+})\\
		&\leq (\nu_y-(p-1)\delta)^+\log(B)+o(\log(B))
	\end{align}
	Recall that ${\bf Y}_{1:p}$ is an invertible function of $(\overline{\bf Y}_{1:p})^{\nu_y}$. The key step \eqref{eq:figjustify} is explained by  Figure \ref{fig:gdof} which shows that $(\overline{\bf Y}_{1:p})^{\nu_y}$  is in turn a function (up to bounded distortion which is inconsequential for GDoF) of the top $\nu_y$ digits of $\overline{\bf X}_0$, the top $(\nu_y-\delta)^+$ digits of $\overline{\bf X}_1$, the top $(\nu_y-2\delta)^+$ digits of $\overline{\bf X}_2$,$\cdots$, and the top $(\nu_y-(p-1)\delta)^+$ digits of $\overline{\bf X}_{p-1}$. Rigorous derivations of such bounds, while tedious, may be found in several recent works \cite{Arash_Jafar_sumset}, so let us omit the  details here.
	Combined with \eqref{eq:GDoF} this gives us our other desired bound: 
	\begin{align}
	\nu&\leq (\nu_y-(p-1)\delta)^+.
	\end{align}

	\bibliography{Thesis}

\begin{thebibliography}{10}
\providecommand{\url}[1]{#1}
\csname url@samestyle\endcsname
\providecommand{\newblock}{\relax}
\providecommand{\bibinfo}[2]{#2}
\providecommand{\BIBentrySTDinterwordspacing}{\spaceskip=0pt\relax}
\providecommand{\BIBentryALTinterwordstretchfactor}{4}
\providecommand{\BIBentryALTinterwordspacing}{\spaceskip=\fontdimen2\font plus
\BIBentryALTinterwordstretchfactor\fontdimen3\font minus
  \fontdimen4\font\relax}
\providecommand{\BIBforeignlanguage}[2]{{%
\expandafter\ifx\csname l@#1\endcsname\relax
\typeout{** WARNING: IEEEtran.bst: No hyphenation pattern has been}%
\typeout{** loaded for the language `#1'. Using the pattern for}%
\typeout{** the default language instead.}%
\else
\language=\csname l@#1\endcsname
\fi
#2}}
\providecommand{\BIBdecl}{\relax}
\BIBdecl

\bibitem{Yu_Maddah-Ali_Avestimehr_Polynomial}
Q.~Yu, M.~A. Maddah-Ali, and A.~S. Avestimehr, ``Polynomial codes: an optimal
  design for high-dimensional coded matrix multiplication,'' in
  \emph{Proceedings of the 31st International Conference on Neural Information
  Processing Systems}, 2017, pp. 4406--4416.

\bibitem{Dutta_Fahim_Haddadpour}
S.~Dutta, M.~Fahim, F.~Haddadpour, H.~Jeong, V.~Cadambe, and P.~Grover, ``On
  the optimal recovery threshold of coded matrix multiplication,'' \emph{IEEE
  Transactions on Information Theory}, vol.~66, no.~1, pp. 278--301, 2019.

\bibitem{GPolyDot}
S.~Dutta, Z.~Bai, H.~Jeong, T.~Low, and P.~Grover, ``{A Unified Coded Deep
  Neural Network Training Strategy Based on Generalized PolyDot Codes for
  Matrix Multiplication},'' \emph{ArXiv:1811.1075}, Nov. 2018.

\bibitem{Yu_Maddah-Ali_Avestimehr}
Q.~Yu, M.~A. Maddah-Ali, and A.~S. Avestimehr, ``Straggler mitigation in
  distributed matrix multiplication: Fundamental limits and optimal coding,''
  \emph{IEEE Transactions on Information Theory}, vol.~66, no.~3, pp.
  1920--1933, 2020.

\bibitem{Yu_Lagrange}
Q.~Yu, S.~Li, N.~Raviv, S.~M.~M. Kalan, M.~Soltanolkotabi, and S.~A.
  Avestimehr, ``Lagrange coded computing: Optimal design for resiliency,
  security, and privacy,'' in \emph{The 22nd International Conference on
  Artificial Intelligence and Statistics}.\hskip 1em plus 0.5em minus
  0.4em\relax PMLR, 2019, pp. 1215--1225.

\bibitem{Reisizadeh_Prakash_Pedarsani}
A.~Reisizadeh, S.~Prakash, R.~Pedarsani, and A.~S. Avestimehr, ``Coded
  computation over heterogeneous clusters,'' \emph{IEEE Transactions on
  Information Theory}, vol.~65, no.~7, pp. 4227--4242, 2019.

\bibitem{Lee_Suh_Ramchandran}
K.~Lee, C.~Suh, and K.~Ramchandran, ``High-dimensional coded matrix
  multiplication,'' in \emph{2017 IEEE International Symposium on Information
  Theory (ISIT)}.\hskip 1em plus 0.5em minus 0.4em\relax IEEE, 2017, pp.
  2418--2422.

\bibitem{Lee_Lam_Pedarsani}
K.~Lee, M.~Lam, R.~Pedarsani, D.~Papailiopoulos, and K.~Ramchandran, ``Speeding
  up distributed machine learning using codes,'' \emph{IEEE Transactions on
  Information Theory}, vol.~64, no.~3, pp. 1514--1529, 2017.

\bibitem{Dutta_Cadambe_Short}
S.~Dutta, V.~Cadambe, and P.~Grover, ``Short-dot: Computing large linear
  transforms distributedly using coded short dot products,'' in \emph{Advances
  In Neural Information Processing Systems}, 2016, pp. 2100--2108.

\bibitem{Dutta_Cadambe_Codedconv}
------, ``Coded convolution for parallel and distributed computing within a
  deadline,'' \emph{arXiv preprint arXiv:1705.03875}, 2017.

\bibitem{Yu_Maddah-Ali_CodedDFT}
Q.~Yu, M.~A. Maddah-Ali, and A.~S. Avestimehr, ``Coded fourier transform,''
  \emph{arXiv preprint arXiv:1710.06471}, 2017.

\bibitem{Jahani-Nezhad_Maddah-Ali}
T.~Jahani-Nezhad and M.~A. Maddah-Ali, ``Codedsketch: A coding scheme for
  distributed computation of approximated matrix multiplications,'' \emph{arXiv
  preprint arXiv:1812.10460}, 2018.

\bibitem{Baharav_Lee_Ocal}
T.~Baharav, K.~Lee, O.~Ocal, and K.~Ramchandran, ``Straggler-proofing
  massive-scale distributed matrix multiplication with d-dimensional product
  codes,'' in \emph{2018 IEEE International Symposium on Information Theory
  (ISIT)}.\hskip 1em plus 0.5em minus 0.4em\relax IEEE, 2018, pp. 1993--1997.

\bibitem{Suh_Lee_Msparse}
G.~Suh, K.~Lee, and C.~Suh, ``Matrix sparsification for coded matrix
  multiplication,'' in \emph{2017 55th Annual Allerton Conference on
  Communication, Control, and Computing (Allerton)}.\hskip 1em plus 0.5em minus
  0.4em\relax IEEE, 2017, pp. 1271--1278.

\bibitem{Wang_Liu_CLT}
S.~Wang, J.~Liu, N.~Shroff, and P.~Yang, ``Fundamental limits of coded linear
  transform,'' \emph{arXiv preprint arXiv:1804.09791}, 2018.

\bibitem{Mallick_Chaudhari_Joshi}
A.~Mallick, M.~Chaudhari, U.~Sheth, G.~Palanikumar, and G.~Joshi, ``Rateless
  codes for near-perfect load balancing in distributed matrix-vector
  multiplication,'' \emph{Proceedings of the ACM on Measurement and Analysis of
  Computing Systems}, vol.~3, no.~3, pp. 1--40, 2019.

\bibitem{Wang_Liu_Sparse}
S.~Wang, J.~Liu, and N.~Shroff, ``Coded sparse matrix multiplication,'' in
  \emph{International Conference on Machine Learning}.\hskip 1em plus 0.5em
  minus 0.4em\relax PMLR, 2018, pp. 5152--5160.

\bibitem{Severinson_iAmat_Rosnes}
A.~Severinson, A.~G. i~Amat, and E.~Rosnes, ``Block-diagonal and lt codes for
  distributed computing with straggling servers,'' \emph{IEEE Transactions on
  Communications}, vol.~67, no.~3, pp. 1739--1753, 2018.

\bibitem{Haddadpour_Cadambe_Finite}
F.~Haddadpour and V.~R. Cadambe, ``Codes for distributed finite alphabet
  matrix-vector multiplication,'' in \emph{2018 IEEE International Symposium on
  Information Theory (ISIT)}.\hskip 1em plus 0.5em minus 0.4em\relax IEEE,
  2018, pp. 1625--1629.

\bibitem{Sheth_Dutta_Chaudhari}
U.~Sheth, S.~Dutta, M.~Chaudhari, H.~Jeong, Y.~Yang, J.~Kohonen, T.~Roos, and
  P.~Grover, ``An application of storage-optimal matdot codes for coded matrix
  multiplication: Fast k-nearest neighbors estimation,'' in \emph{2018 IEEE
  International Conference on Big Data (Big Data)}.\hskip 1em plus 0.5em minus
  0.4em\relax IEEE, 2018, pp. 1113--1120.

\bibitem{Jeong_Ye_Grover}
H.~Jeong, F.~Ye, and P.~Grover, ``Locally recoverable coded matrix
  multiplication,'' in \emph{2018 56th Annual Allerton Conference on
  Communication, Control, and Computing (Allerton)}.\hskip 1em plus 0.5em minus
  0.4em\relax IEEE, 2018, pp. 715--722.

\bibitem{Kim_Sohn_Moon_Group}
M.~Kim, J.-y. Sohn, and J.~Moon, ``Coded matrix multiplication on a group-based
  model,'' \emph{arXiv preprint arXiv:1901.05162}, 2019.

\bibitem{Park_Lee_Sohn}
H.~Park, K.~Lee, J.-y. Sohn, C.~Suh, and J.~Moon, ``Hierarchical coding for
  distributed computing,'' \emph{arXiv preprint arXiv:1801.04686}, 2018.

\bibitem{Li_Maddah-Ali_Fog}
S.~Li, M.~A. Maddah-Ali, and A.~S. Avestimehr, ``Coding for distributed fog
  computing,'' \emph{IEEE Communications Magazine}, vol.~55, no.~4, pp. 34--40,
  2017.

\bibitem{Jia_Jafar_CDBC}
Z.~Jia and S.~A. Jafar, ``Cross subspace alignment codes for coded distributed
  batch computation,'' \emph{IEEE Transactions on Information Theory}, vol.~67,
  no.~5, pp. 2821--2846, 2021.

\bibitem{Chen_Jia_Wang_Jafar}
Z.~{Chen}, Z.~{Jia}, Z.~{Wang}, and S.~A. {Jafar}, ``Gcsa codes with noise
  alignment for secure coded multi-party batch matrix multiplication,''
  \emph{IEEE Journal on Selected Areas in Information Theory}, vol.~2, no.~1,
  pp. 306--316, 2021.

\bibitem{Fahim_Cadambe}
M.~Fahim and V.~R. Cadambe, ``Numerically stable polynomially coded
  computing,'' \emph{IEEE Transactions on Information Theory}, vol.~67, no.~5,
  pp. 2758--2785, 2021.

\bibitem{Ramamoorthy_Li}
A.~Ramamoorthy and L.~Tang, ``Numerically stable coded matrix computations via
  circulant and rotation matrix embeddings,'' \emph{arXiv preprint
  arXiv:1910.06515}, 2019.

\bibitem{Subramaniam_random}
A.~M. Subramaniam, A.~Heidarzadeh, and K.~R. Narayanan, ``Random
  khatri-rao-product codes for numerically-stable distributed matrix
  multiplication,'' in \emph{2019 57th Annual Allerton Conference on
  Communication, Control, and Computing (Allerton)}.\hskip 1em plus 0.5em minus
  0.4em\relax IEEE, 2019, pp. 253--259.

\bibitem{Gautschi_Vdm}
W.~Gautschi and G.~Inglese, ``Lower bounds for the condition number of
  vandermonde matrices,'' \emph{Numerische Mathematik}, vol.~52, no.~3, pp.
  241--250, 1987.

\bibitem{Pan_Vandermonde}
V.~Y. Pan, ``How bad are vandermonde matrices?'' \emph{SIAM Journal on Matrix
  Analysis and Applications}, vol.~37, no.~2, pp. 676--694, 2016.

\bibitem{Cadambe_AMatdot}
H.~{Jeong}, A.~{Devulapalli}, V.~R. {Cadambe}, and F.~{Calmon},
  ``{$\epsilon$-Approximate Coded Matrix Multiplication is Nearly Twice as
  Efficient as Exact Multiplication},'' \emph{arXiv e-prints}, p.
  arXiv:2105.01973, May 2021.

\bibitem{Etkin_Tse_Wang}
R.~Etkin, D.~Tse, and H.~Wang, ``{Gaussian interference channel capacity to
  within one bit},'' \emph{IEEE Transactions on Information Theory}, vol.~54,
  no.~12, pp. 5534--5562, 2008.

\bibitem{Arash_Bofeng_Jafar_BC}
A.~{Gholami Davoodi}, B.~Yuan, and S.~A. Jafar, ``{GDoF} region of the {MISO
  BC}: Bridging the gap between finite precision and perfect {CSIT},''
  \emph{IEEE Transactions on Information Theory}, vol.~64, no.~11, pp.
  7208--7217, Nov. 2018.

\bibitem{Motahari_Gharan_Khandani_real}
A.~Motahari, S.~{Oveis Gharan}, and A.~Khandani, ``Real interference alignment
  with real numbers,'' Aug 2009, arXiv:0908.1208.

\bibitem{Arash_Jafar}
A.~G. Davoodi and S.~A. Jafar, ``{Aligned image sets under channel uncertainty:
  Settling conjectures on the collapse of degrees of freedom under finite
  precision CSIT},'' \emph{IEEE Transactions on Information Theory}, vol.~62,
  no.~10, pp. 5603--5618, 2016.

\bibitem{Cadambe_Jafar_int}
V.~Cadambe and S.~Jafar, ``{Interference Alignment and the Degrees of Freedom
  of the $K$ user Interference Channel},'' \emph{IEEE Transactions on
  Information Theory}, vol.~54, no.~8, pp. 3425--3441, Aug. 2008.

\bibitem{Jafar_FnT}
S.~Jafar, ``{Interference Alignment: A New Look at Signal Dimensions in a
  Communication Network},'' in \emph{Foundations and Trends in Communication
  and Information Theory}, 2011, pp. 1--136.

\bibitem{Arash_Jafar_sumset}
A.~{Gholami Davoodi} and S.~A. {Jafar}, ``Sum-set inequalities from aligned
  image sets: Instruments for robust {GDoF} bounds,'' \emph{IEEE Transactions
  on Information Theory}, vol.~66, no.~10, pp. 6458--6487, 2020.

\end{thebibliography}
	\bibliographystyle{IEEEtran}
\end{document}